\documentclass{aastex}

\setcitestyle{citesep={;}}
\def\ee#1{\ifmmode {} \times 10^{#1} \else ${} \times 10^{#1}$\fi}

\def\aboutless{$\simless$\kern.03em}

\def\aboutmore{$\simgreat$\kern.03em}

\begin{document}

\title{\bf {CO in distantly active comets}}

\bigskip

\bigskip

\author{M. Womack$^1$, G. Sarid$^2$, K. Wierzchos$^1$}

\affil{$^1$University of South Florida, $^2$Florida Space Institute, University of Central Florida}

\section{Abstract}

The activity of most comets near the Sun is dominated by the sublimation of frozen water, the most abundant ice in comets. Some comets, however, are active well beyond the water-ice sublimation limit of $\sim$ 3 AU.  Three bodies dominate the observational record and modeling efforts for distantly active comets: the long-period comet C/1995 O1 Hale-Bopp and the short-period comets (with Centaur orbits) 29P/Schwassmann Wachmann 1 and 2060 Chiron. We summarize what is known about these three objects with an emphasis on their gaseous comae. We calculate their CN/CO and CO$_2$/CO production rate ratios from the literature and discuss implications. Using our own data we derive CO production rates, Q(CO), for all three objects, in order to examine whether there is a correlation between gas production and different orbital histories and/or size. We further examine the applicability of existing models in explaining the systematic behavior of our small sample. We find that orbital history does not appear to play a significant role in explaining 29P's CO production rates. 29P outproduces Hale-Bopp at the same heliocentric distance, even though it has been subjected to much more solar heating. Previous modeling work on such objects predicts that 29P should have been devolatilized over a fresher comet like Hale-Bopp.  This may point to 29P having a different orbital history than current models predict, with its current orbit acquired more recently. On the other hand, Chiron's CO measurements are consistent with it being significantly depleted over its original state, perhaps due to increased radiogenic heating made possible by its much larger size or its higher processing due to orbital history. Observed spectral line profiles are consistent with the development and sublimation of icy grains in the coma at about 5-6 AU for 29P and Hale-Bopp, and this is probably a common feature in distantly active comets, and an important source of volatiles for all comets within 5 AU. In contrast, the narrow CO line profiles indicate a nuclear, and not extended, origin for CO beyond $\sim$ 4 AU.


\section{Introduction}

As a typical comet approaches the inner solar system it develops a coma 
when frozen water begins to sublimate at
$\sim$ 3 AU from the Sun.\footnote{This is the heliocentric distance at which 
the sublimation temperature of water ice, T$_{sub} \sim$ 150 
K, assuming a gas density of $\sim$ 10$^{-13}$ cm$^{-3}$, \citep{yam85}, equals 
the blackbody temperature of a fast rotating spherical comet nucleus, $T_{bb}$, according to
$$T_{bb} = \left({{1-A}\over{\epsilon} }\right)^{1/4}T_{\odot}\left({ 
{R_{\odot}}\over{2r} }\right)^{1/2}, \eqno(1) $$
\noindent where A = nucleus bond albedo (typically assumed to be
in the range of 0.03 -- 0.15), $\epsilon$ is the surface emissivity 
($\sim$ 0.5 - 0.9), R$_{\odot}$ and T$_{\odot}$ 
are radius and temperature of the Sun, respectively.} 
Some comets, however, exhibit measurable
activity beyond this water ice sublimation boundary \citep{whi55,roe62,mee87,hug91,sek92},
and up to a third of all comets observed become active 
beyond 3 AU \citep{epifani07}. Sublimation of water ice may still generate some activity beyond
the normal water-ice sublimation boundary of $\sim$ 3 AU. Even at very large distances, 
water ice sublimation can still occur if triggered by impacts from other small bodies, tidal 
disruption from interactions with massive planets, and interactions with solar flares and
solar wind erosion \citep{str83,ste95,sek94,bos94}. Collectively, however,
these possible water-ice sublimation triggers are unlikely to generate the long-lived 
comae observed in most distantly active comets \citep{mum93}. Thus, most
comets that are active beyond $\sim$ 3 to 4 AU should be considered to have comae which 
are generated by a volatile other than water.

Understanding how comae are generated in comets is critical to developing accurate models
of nucleus composition. The activity of distant comets also has important implications for other
icy objects in the outer solar system, some of which may share common origins
with comets. For example, some short-period comets may be 
collision fragments from Kuiper Belt Objects (KBOs) \citep{far96,dur00,schul02,done15}.
Also, KBOs, Pluto and even icy moons may also experience 
similar physical and chemical conditions as distantly active comets \citep{wei99,boc01}.

Analysis of visible and infrared observations often provide useful information, such as
the extended comae of distant comets are composed primarily of dust
grains with a total mass of $\sim$ 10$^{9}$kg and coma expansion velocities 
of 0.3 km s$^{-1}$ or less \citep[also see ][]{whi55,hug91, trigo10,kuly16}. Visible lightcurves 
often show variations of two to four magnitudes 
over a few weeks and much smaller amplitude changes over a few hours. The larger amplitude 
changes are typically assumed to be due to significant mass loss; on the other hand, smaller amplitude variations are 
often attributed to the rotation state of the comet's nucleus \citep{whi80,mee93}.



In addition to dust, gaseous emission has also been detected. 
CO$^+$, CN and OH were the first volatiles observed in distantly active comets 
\citep{coch80, wyc85}, but they cannot drive the activity. This is because they are produced 
through ionization or dissociation of other molecules in the coma and do not exist in the nucleus. Observations of
these species, however, are still useful as tracers of their parent molecules 
\citep{coch91,AHearn84,feld04}. For example, spectra of 
OH emission may be useful in determining whether H$_2$O (the presumed parent of OH) 
is sublimating. When CN is observed, then HCN is likely to be 
abundant near the surface or in the coma. However, we cannot rule out a distributed
source, such as C$_2$N$_2$, HC$_3$N or even N-bearing refractories as a parent of CN 
\citep{feld04, desp05, fes98}. CO is difficult to photoionize
beyond 5 AU from the Sun, so detecting CO$^+$ implies either unusual ionization of 
CO, or more likely, a very large amount of CO being released in the coma. Thus, CO$^+$
may be a useful tracer of CO emission \citep{magn86,coch91}.

Carbon monoxide is of particular interest, because of
its identification as a ``supervolatile" (cosmogonically abundant molecules with sufficient 
equilibrium vapor pressures to vigorously sublimate at very large ($>$ 35 AU) heliocentric 
distances) (see Table \ref{tab:svolatile}). A breakthrough in understanding distantly active comets occurred
when CO gaseous emission 
was detected at millimeter wavelengths 
in comet 29P/Schwassmann-Wachmann 1 (hereafter referred to as 29P) 
at $\sim$ 6 AU in amounts significantly higher than the dust production 
rates \citep{sen94}.  Carbon monoxide was already suspected of being in 29P's coma, 
because of the previously observed CO$^+$ in 29P \citep{lar80,coch91} and the very low sublimation 
temperature of CO ice; indeed, its detection in comets was predicted based on favorable excitation conditions 
and expected production rates \citep{cro83}. The following year, CO emission was also 
detected in C/1995 O1 (Hale-Bopp) over 6.8 - 6.0 AU \citep{jew96,biv96,wom97}, and 
in 2060 Chiron at 8.5 AU \citep{wom99}. Doubts were raised about the Chiron CO detection 
\citep{boc01}; however, an independent re-analysis \citep{jew08} maintained that the 
CO line was formally detected in Chiron. The derived CO production rates 
were high enough that CO outgassing was assumed to be the main cause of activity for all three
objects (Table \ref{tab:volatile}). 

Other highly volatile and abundant gases, such
as CO$_2$, CH$_4$, and N$_2$ may contribute to distant activity, but are 
difficult to observe \citep{pri04,oot12,rea13,bau15}. 
CO$_2$, CH$_4$ and N$_2$ do not possess an electric 
dipole moment and thus have no rotational transitions, which are needed for observations at
millimeter wavelengths. They have transitions at infrared wavelengths, but 
the emission often suffers telluric contamination and are weakly excited by solar radiation beyond 4 AU. 
Observations from space-based telescopes indicate that CO$_2$ is 
much more abundant than CO in a large survey of distant comets (see summaries by \citet{oot12,mckay12}). 
CO$_2$-dominated outflows are documented 
in comets 103P/Hartley2 \citep{AHearn11} and 67P from spacecraft data \citep{hassig15}, which show that CO$_2$ is a major player
for these comets closer to the Sun. As we discuss later, all available data indicate that CO$_2$ is not likely 
responsible for most of the observed activity in 29P, Hale-Bopp, and Chiron. Methane has not been detected in distant 
comets, although significant limits were set for 29P (see Table \ref{tab:volatile}). N$_2$ emission
in comets can be estimated from optical spectra of N$_2^+$ \citep{wom92,lutz93,coch00,
coch02,iva16} and with mass spectroscopy from
spacecraft data \citep{Rub15}. All detections, or non-detection limits, point to 
N$_2$ abundances being much lower than that of CO in comets,
including Hale-Bopp and 29P. In addition, the nitrogen 
depletion issue in comets \citep{coch00,iro03} make it unlikely that N$_2$ plays a substantial role in
driving distant comet activity.

Other volatiles have been detected beyond 3 AU, but their low production 
rates also indicate that they are unlikely to generate much of the 
observed distant activity. For example, CH$_3$OH, HCN, H$_2$CO, H$_2$S and CS were identified 
in spectra of Hale-Bopp far from the Sun, but with production rates all significantly
lower than CO production rates \citep{biv97,wom97}. 


Although most cometary studies focus on abundances from near comets, large heliocentric coverage is 
also needed to aid compositional studies of all comets \citep{dellorusso16}. 
Although many measurements exist of distant comets, the causes
of the activity are not well understood. In this paper we
summarize possible energy sources and observational results, and make a comparative study of 
29P, Hale-Bopp and Chiron. Although the dust and gas counterparts 
of the comae are both addressed, the main emphasis is on the volatiles (mainly CO), 
since they are directly involved in generating the distant activity. Studying CO's behavior beyond 4 AU is especially
important to confirm its natal contributions and study it in isolation \citep{pierce10}.

\section{Proposed Mechanisms to Generate Comae Far from the Sun}

Solar radiation is the most significant energy source for comets. Consequently, 
the heliocentric distance, albedo and seasonal effects of the nucleus 
all play vital roles in how the energy is balanced in a comet on the surface 
and in the sub-surface layers of a comet nucleus. Here we review energy 
sources that may play a significant role in sustained activity at large distances, such as 
supervolatile sublimation, crystallization of amorphous water ice, and radioactive decay.

First, we consider supervolatile sublimation and outgassing. If comet material started 
off very cold (such as the 
temperatures needed initially to maintain CO, N$_2$ or CH$_4$ in solid form), 
then their deep interiors may not have experienced any warming episodes, and 
highly volatile species can be preserved in the nucleus. Sublimation of such
``supervolatile" ices can drive activity far from the Sun, and possibly 
beyond 40 AU for the most volatile CO. One concern raised about 
supervolatile ices is that severe devolatilization may 
have occurred due to significant heating over the $\sim$ 4.6 Gyr since the comets formed. 
As \citet{kou01} point out, however, laboratory results 
indicate an extremely low thermal conductivity for water ice in
comets, suggesting that heating is negligible below the outer several tens of
centimeters, and so extends the lifetime of supervolatiles in comet nuclei. Additionally, we 
point out the abundant evidence for cometary supervolatiles, as 
recorded by observations and space mission experiements, which show that supervolatiles
are long-lived in cometary nuclei and have not been exhausted close to the Sun 
or preferentially in comets that sustain more heating (such as Jupiter Family comets, which have 
low orbital inclinations and 
orbital periods shorter than $\sim$ 20 years) as compared to new comets \citep{AHearn11,
ahearn12,hassig15,dellorusso16}.

Next, we examine the amorphous water ice crystallization phase change in comet nuclei. 
If cometary water originates primarily from preserved interstellar grains, or from the 
distant outer regions of the solar disk, then it is likely to accumulate in an amorphous 
state \citep{kou94,irv00}. Upon heating, such ice undergoes a solid phase transition to 
a crystalline structure. The lab-measured threshold for transition to cubic ice is 
$\sim$100 K, with another transition from cubic to hexagonal ice at $\sim$160 K 
\citep{lauf87}. However, only the first transition is exothermic and it involves 
a significant change in specific density and arrangement of ice structure, so that 
trapped molecules can be released. The functional form for the amorphous-crystalline 
transition, with 95\% of the ice changing to the crystalline phase in a time (sec) 
$t_c$, was derived as
$$t_c=9.54\ee{-14}\exp(5370/T) \eqno(3) $$
\noindent at a temperature T (K), for the range of 125-150 K, \citep{schm89}. Extrapolation 
to lower temperatures gives 120 K as the activation temperature, where the rate becomes rapid 
($\sim$1 month). This is commonly used as a nominal crystallization temperature. 
 
This phase transition could release enough energy, approximately 
10$^9$ ergs g$^{-1}$ \citep{gho68, Kli81}, to increase the temperature 
of the surrounding ice by another 45 K \citep{enz97}. This, in turn,
could trigger even more crystallization and sublimation, 
creating a runaway crystallization effect \citep{pat74,pri92}. 
Accordingly, much of the outermost ice of a comet that has been heated 
to $\sim$ 140 K is expected to be in the crystalline form. This could
mean that most, if not all, of the comets that make repeated approaches 
within 5 AU \footnote{Four to five AU is the approximate heliocentric range where a nucleus
is expected to reach 140K due to solar heating, see equation 1.}
 of the Sun should have crystalline water ice on the outermost layers of the nucleus.

In addition to releasing energy, this phase transition decreases the
density of ice from $\rho_a$ 
$\sim$ 1.2 g cm$^{-3}$ to $\rho_c$ $\sim$ 0.9 g cm$^{-3}$. This structural
change releases gas molecules that were 
trapped in the original ice, such as CO and CO$_2$ \citep{bar88,bar95,pri95}.

Runaway amorphous ice crystallization is modulated by the fact that laboratory 
studies show that exothermic crystallization occurs only 
for pure amorphous ice. Slight impurity levels (such as $>$ 2\% in mol of
CO in the ice matrix) in the ice are enough to reduce the effective latent heat of crystallization, 
thus reducing the exothermic nature of the process \citep{kou01}. This could result in 
isolated bursts, rather than crystallizing all at once. Crystallization of amorphous water ice 
is a mechanism included in many theoretical models of distantly active comets \citep[e.g., ][]{pri90,prialnikbarnun90, 
cap00,sar05}.
While a few other exothermic mechanisms have been suggested, such as polymerization of condensed-phase 
HCN \citep[e.g., ][]{ret92,mat95} and melting and dissolution of gases \citep{milesheat16}, these processes 
have not been put to a quantitative test or included in a full account of the energetics involved in 
cometary evolution.

Radiogenic heating is an additional source of energy in distant comets that may 
deplete significant amounts of frozen CO and other supervolatiles from the nucleus 
\citep{pri87,har93,pri95,des01}. The effectiveness of this mechanism depends critically on 
such parameters as the diffusivity and thermal conductivity of the nucleus, both of which have a large range in values 
for the ice and silicate components. However, if the nucleus is
smaller than a radius of $\sim$10 km, then the cooling timescale from 
radiogenic heating is shorter 
than the heating timescale. These objects would effectively get rid of the excess 
energy generated by radioactive decay quicker than this energy can promote a thermodynamic
heating cycle on icy layers \citep{pria95, pri08}. 
Hence, for comets with nuclei smaller than 10km, the initial temperature is probably preserved deeper 
than about 100 meters.

One important concern with modeling cometary ice is that the experimental value 
of the thermal conductivity of water ice is 10$^4$ times lower than the theoretical value \citep{kli80,kou92}, yet both quantities have been used in models of distantly active comets. The grain packing configuration and appreciable porosity can change the thermal 
conductivity by orders of magnitude. It is affected by the porosity, Hertz factor
(the contact area relative to cross-sectional area of material grains) and the specific thermal conductivity of the ice (what phase of ice and its level of impurity) and pores (depends on the shape and inter-connectivity of the porous space). The surface-to-volume ratio of the pores is also sensitive to the actual pore size distribution, at any given layer of the nucleus, and can vary by a few orders of magnitude \citep{sar05}. 

If CO sublimation generates activity in a distant comet, a simple estimate of the 
sublimating area, $a$, on the nucleus can be derived using the following equation, assuming 
that the nucleus is a rapid rotator 
$$a = { {(dM/dt)r^2L_{CO}}\over{F_{\odot}(1-A)} }, \eqno(4) $$

\noindent where $dM/dt$ is the total mass loss rate of CO, $F_{\odot}$ = 1360 W/m$^2$, $r$ is 
heliocentric distance in AU, $L_{CO}$ is the latent heat of sublimation for
CO, and $A$ is the Bond albedo of the nucleus. For example, using dM/dt = 2000 kg s$^{-1}$
from the CO observations, this estimate gives a circular spot 
with an emitting radius of $\sim$ 2 - 6 km, depending on whether the CO
is released as a sublimating ice ($L_{CO}$ $\sim$ 4\ee{5} J/kg), 
or as a trapped gas in water ice ($L_{CO} \sim$ 1.5\ee{6} J/kg) \citep{bro79}, corresponding to 
$<$ 0.1\% of the surface area for 29P and Hale-Bopp.

As a cautionary statement, one should consider that sometimes volatiles detected in the coma did not originate 
directly from the comet's nucleus, and thus do not play a role in generating activity. 
For example, a source could be ice-coated grains of refractory material
expelled from the nucleus \citep{del71,han81,han84}. Once in the solar 
radiation field, there is enough energy available 
(if within $\sim$ 5-6 AU of the Sun) for the ice-coatings to sublimate.
Such grains reportedly are destroyed very quickly (within 20 hours for 1 micron 
sized grains) once their temperatures reach $\sim$ 150K \citep{lam74}.

Evidence for the sublimation of icy grains was seen 
in the emission of OH at 308 nm in comet Bowell, which was detected as 
the comet moved from 5 to 3 AU  
\citep{AHearn84,ahearn95}. 
Icy grains with radii of 7 - 30 micron were determined to exist in Hale-Bopp's coma with a 
median lifetime of about two days at 2.9 AU \citep{lel98}. Further evidence is found from spectral 
line-widths of methanol and other species in Hale-Bopp at $\sim$ 5 AU, which were much
wider than one would expect for molecules sublimating directly from the nucleus 
\citep{wom97}. 103P/Hartley 2 was also observed by the EPOXI mission and the large
effective sublimation area derived from photometric observations was determined 
to be from ice grains sublimating, but at much closer distances to the Sun \citep{proto14, kel13}. 


\section{Observations of Distantly Active Comets}

In this section, we review what is known about each object, and in the next section we examine the similarities and 
differences between these objects and propose a common model that may explain the distant activity.

\subsection{Comet Hale-Bopp}

Comet Hale-Bopp presented an exceptional opportunity to study distant
activity, since it already had a giant coma upon discovery at 7 AU. 
It is a long-period comet with a high-inclination orbit, and, thus 
has a very different orbital history than either 29P or Chiron. Because it has spent most of its lifetime very
far from the Sun, it is probably less processed than many other comets. 
It is a very large comet - its nucleus radius was measured to be in the range of $R$ = 20 - 35 km 
\citep{wea99,sek97,fer99}, and $R_{Hale-Bopp}$ = 30 $\pm$ 10 km was recommended as a compromise
\citep{fer02}, which we adopt for calculations in a later section. Hale-Bopp's radius is very similar to 29P, 
but much smaller than Chiron. 

Hale-Bopp exhibited several changes as it approached the Sun.
There were reports of a swirled ``jet" and a few outbursts in the dust coma 
from 7 to 6 AU  \citep{sek96,weaver97}. Multiple porcupine-like 
features were first observed during May-Jul 1996 at $r \sim$ 4 AU and 
continued for several months during perihelion approach \citep{boe99,bra99,weaver97}.
Surface brightness profiles of Hale-Bopp's coma at $\sim$ 7 AU were fit with
a power law and had a slope of $s$ =  $-0.9$ to $-1.2$ \citep{weaver97}, which is 
consistent with a spherically symmetric outflowing coma 
subject to solar radiation pressure. 

Numerous records were set for the  
detections of cometary gases at large heliocentric distances with Hale-Bopp, 
including the first time that CH$_3$OH, HCN, H$_2$S, CO$_2$ and H$_2$CO were 
observed in a comet beyond 4 AU  \citep{wom97,biv97b,woo97,cro96}. Many of these 
species were observed at even larger distances post-perihelion 
\citep{biv97b,biv02}. Carbon monoxide emission was first detected in Hale-Bopp 
at $\sim$ 6.8 AU \citep{jew96,biv96} with production rates high enough to 
cause the observed dust production. CO$_2$ emission was detected at 4-5 AU via its vibrational
bands at 4.3 $\mu$m and inferred from CO Cameron band emission, 
present at $\sim$ 15\% of the production rate of CO 
\citep{cro99,weaver97}. A strong upper limit was obtained for N$_2^+$ in Hale-Bopp near perihelion by \citet{coch00}, 
which when compared with CO$^+$ measurements is also consistent with N$_2$ produced
in much smaller quantities than CO. The production rates of OH (a proxy for H$_2$O), 
CH$_3$OH, HCN, H$_2$S and H$_2$CO were all much lower than those of CO. We conclude 
that Hale-Bopp's gas coma was dominated by CO whenever the comet was beyond 4 AU pre-perihelion \citep{weaver97,wom97,biv02}.

The physical state of Hale-Bopp's coma volatiles is revealed by their millimeter-wavelength 
spectral line profiles. 
For example, CO spectra had a velocity blue-shift of $\sim$ 0.4 km s$^{-1}$, and
narrow line-widths of FWHM $\sim$ 0.3-0.4 km s$^{-1}$ \citep{biv96,jew96}. 
The narrow line-widths are consistent with originating from 
deeper and colder regions in the nucleus. At larger distances, the CO line profile occasionally 
exhibited two narrow blue-shifted peaks \citep{wom97,gun03}; however,
a single peak was observed most of the time. The velocity vector component of the spectra shows
a sunward release of CO into the coma. The CO line-shape was 
different post-perihelion and exhibited a red-shifted peak along with a blue-shifted peak.
The double peaks were interpreted by \citet{gun03} as evidence for both nuclear and 
extended CO sources, similar to their model for 29P. 

\begin{figure}

\epsscale{.6}

\plotone{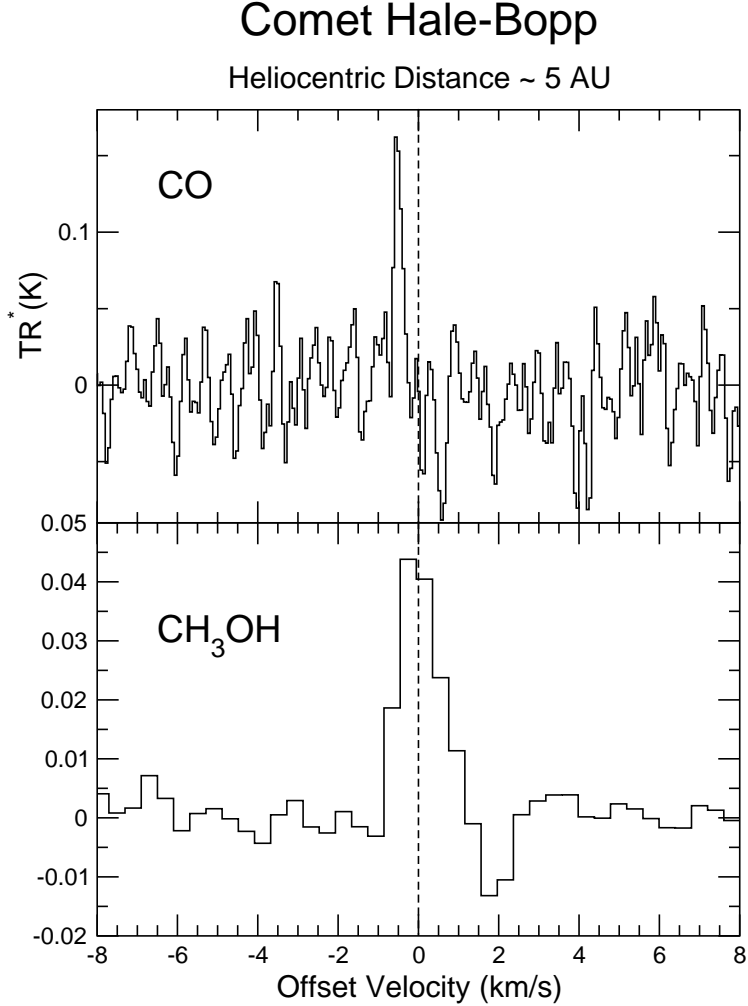}

\caption{
\label{fig:cometh}
Spectra of CO and CH$_3$OH emission from comet C/1995 O1 Hale-Bopp at ~5 AU, pre-perihelion, obtained with the NRAO 12-m
telescope. The CO J=2-1 spectrum
was obtained at 230 GHz on 1996 March 31 with 49 kHz/channel resolution, which corresponds to 0.06 km/s/channel; 
the methanol emission was obtained at 145 GHz on 1996 Apr 16 with 195 kHz/channel resolution, which corresponds to 0.40 km/s/channel.
The dotted line indicates the comet's rest frame velocity. As the figure shows, the CO emission is blue-shifted from the ephemeris
velocity of the comet and has a very narrow line-width. In contrast, 
the methanol emission has no measurable shift from the comet's velocity
and has a broader line-width. This difference can be explained if CO
emission originates from within the cold comet nucleus and the methanol
escapes from hot grains in the coma.
}
\end{figure}

Interestingly, beyond $\sim$ 4 AU, the spectral line profiles of most other 
species were significantly different from that of CO. For example, CH$_3$OH, 
OH, HCN and H$_2$S had much broader line-widths than CO, and very little, if 
any, velocity shifts (see Figure \ref{fig:cometh}). Their line-widths were 
FWHM $\sim$ 0.7 to 1.5 km s$^{-1}$ over the range of 5 to 4 
AU. Beyond 4 AU, CO linewidths were typically much narrower, 0.4 km s$^{-1}$, 
although occasionally the line was observed to be much broader \citep{wom97,biv02}. 
Since the spectral linewidths are mostly a result of the thermal distribution of the gas, 
different production mechanisms between CO and the other molecules are responsible for 
most of the observed variation in line profiles. The broader, cometocentric line profiles may be explained if 
H$_2$O, CH$_3$OH, HCN, H$_2$CO and H$_2$S were first released into
the coma on grains, and then sublimated when they accumulated enough 
energy from the solar radiation field. Volatiles released from 
a symmetric halo surrounding the nucleus will appear centered on the ephemeris
velocity. They will also be at a higher temperature and have a larger velocity 
distribution (and larger line-widths) than those released directly off the 
nucleus. Occasionally, HCN and CH$_3$OH spectra mimicked the CO line profile 
shape, which may indicate that they were produced from two or more methods \citep{gun08}.
Unfortunately, the CO$_2$ band at 4.3 microns in Hale-Bopp is not 
spectrally resolved and, so, one cannot derive a useful upper limit to 
velocity information of the CO$_2$ emission.
 
The pre-perihelion production rates and narrow line profiles of CO, and dust production rates, in Hale-Bopp 
are best reproduced by nucleus models that contain amorphous water ice with trapped CO gas, 
small amounts of CO ice, and crystallization of water ice starting at $\sim$ 7 AU from the Sun 
\citep{pri99,enz99,cap02}.
Prior to crystallization, sublimation of solid CO is invoked as the dominant source of
activity, without detailing the source of CO ice in the nucleus. The models predict slightly 
higher production rates post-perihelion than pre-perihelion, which was not seen by \citet{biv02,gun03}. 
This suggests that thermal inertia was not very important for Hale-Bopp once the comet was beyond 4 AU, and therefore 
most of the activity is at or near the surface of the nucleus. Observations are also consistent with the activity
being well-correlated with insolation, as evidenced by symmetric behavior of gas and dust production rates before and 
after perihelion \citep{biv02}.



\subsection{Chiron}

2060 Chiron is an unusual icy minor planet. When it was discovered in 
1977, it was categorized as the most distant 
known asteroid.  Its reflectance spectrum had a  relatively neutral 
``color",  similar to C-class asteroids \citep{leb84,har90,ster14}, and it occasionally 
outbursts and develops a coma, which warrants it a formal comet designation  
\citep{tho88,mee90}. Evidence for CN and CO gaseous emission \citep{bus91,wom99} 
further secured the cometary identity.

In addition to its cometary designation, orbital dynamical studies place Chiron among the small population of large, outer solar system objects called Centaurs, which
are moderately sized Kuiper Belt Objects that have moved in closer, with 
unstable orbits, and cross one or more giant planet orbits \citep[e.g., ][]{fer80,ste89,hah90,lev94,fer94}. Its brightness varies significantly on both short and long timescales, with 
changes in V and R mag of $\sim$ 0.05 $-$ 0.10 magnitudes over hours and $>$ 1.5 mag over 
months and years \citep{tho88,bus89,luu90,buratti91,duff02}. A water-ice signature was detected in the 
near-infrared, which also appears to vary over time \citep{fos99,luu00,rom03}. It is not known whether 
the ice is primarily crystalline or amorphous; however, the measured color temperature (the equivalent
blackbody temperature that is derived from the continuum spectrum of an object) of the nucleus is
high enough (125 —- 140 K) to trigger the crystallization phase change \citep{cam94}. 

One of the most puzzling aspects of Chiron is that none of these changes are correlated well with 
heliocentric distance. In fact, Chiron reached peak brightness in the visible in 1989
at $\sim$ 11 AU, several years before perihelion, and it may have been even brighter than
this at aphelion, 19 AU \citep{bus93}. For decades, the light curve was largely unexplained - 
other than some of the short-term variations being attributed to reflection off a 
rotating nucleus \citep{bus89,mar93,laz97}. Attempts were made to fit the remaining light curve 
contributions with a model of amorphous water ice and
a highly volatile ice, such as CO and/or CH$_4$, with a dust mantle \citep{fan97}, but they
could not explain the lack of correlation with heliocentric distance. 

Recently, however, these unusual changes in light curve and spectra have been explained as due to 
reflected light from a ring system circling Chiron, based on occultation data \citep{ort15,rup15}.
The proposed rings can account for most of the short- and long-term brightness variations, largely 
by a change in their orbital tilt, if the ring particles have a different albedo than the surface. 
This may also explain the observed changes in the water-ice spectra, provided the ice is 
in the rings and not on the surface. Interestingly, another large Centaur, Chariklo, has a 
ring system, and similar changes are noted there \citep{duff14}.  If confirmed in Chiron, this 
suggests a paradigm shift, where rings may be commonly found around Centaurs and perhaps other 
large icy objects, like Kuiper Belt Objects, and rings need to be included as possible contributors to 
observed differences among KBOs. We note, however, that the Pluto system does not have any rings and the New Horizons project
searched very thoroughly for them. So, while rings could be contributors, for some of our best examples of KBOs,
rings are not present.

The evidence is very strong for a model where most of Chiron's brightness changes are due to the 
changing orbital tilt of a ring system and nucleus rotation; however, these cannot explain the occasional outbursts and development of a coma. 
One possibility is that the rings and coma may even be interdependent, as the most likely formation mechanism for the rings is proposed to be a debris disk fed by some cometary activity, with a possibility that in-falling material from the rings could trigger outbursts \citep{rup15}. If instead of rings, \citet{rup15} also raise 
the possibility that narrow jets may generate activity by a low-gravity analog to the geyser-like vents 
that are sometimes observed on Neptune's large satellite Triton or on Pluto, fueled by sub-surface geological activity. 
More observations are needed of Chiron, in and out of outburst, to confirm the rings or jets.

Chiron's outbursts and coma occur at distances from the Sun too far to be caused by H$_2$O and CO$_2$ ice sublimation, the drivers of most normal comets 
near the Sun. Instead, the outgassing activity is probably largely due to the sublimation of even more volatile species. 
CO is the most likely 
candidate. Its emission in Chiron was singularly detected via the 1-0 rotational line in 1995 with production rates high enough to drive the observed dust coma \citep{wom99,jew08}. The measured CO production rate indicates that much less than 1\% of Chiron's surface was active in 1995, consistent with CO-driven vents or geysers. 

Occultation and thermal emission measurements indicated that Chiron has a very large diameter, ranging from 
D = 150 -- 220 km \citep{cam94,ell95,bus96,fer02}. Following
the discussions by \citet{groussin04} and \citet{fornasier13}, we recommend a diameter of 
D$_{Chiron}$=218$\pm$20 km, which we use for later calculations. Chiron's nucleus is clearly larger than those of Hale-Bopp and 29P, and 
significantly larger than most comet nuclei.

Chiron has a low, but steady dust production rate of $\sim$ 0.5 kg s$^{-1}$, with a mean speed of 
$\sim$ 100 m s$^{-1}$ for a coma that is populated by 1-micron sized grains, according to analysis of 
near-infrared H- and R-band observations \citep{luu90}.  On the other hand, modeling of V-band 
images indicated that the coma was dominated by much larger 100-micron sized dust grains that were largely un-sampled by the near-infrared data \citep{west91,ful94,rom03}. The larger grain sizes lead to a much higher mass loss rate of 20 $\pm$ 10 kg s$^{-1}$, and slower dust expansion velocities of $\sim$  5 to 10 m s$^{-1}$. This is a very low dust production rate when compared to 29P or Hale-Bopp.


Interestingly, Chiron's color temperature measurements are 40 - 50 K warmer than expected for its nucleus if one assumes an equilibrium temperature for a fast rotator spherical blackbody.
It is not known whether the aforementioned ice is primarily crystalline or amorphous; however, the measured color temperature of the nucleus is high enough (125 - 140 K) to trigger the crystallization phase change \citep{cam94}. Unlike 29P, Chiron's nucleus rotation period is well determined to be 5.917813
$\pm$ 0.000007 hours \citep{mar93}. 

Information about Chiron's gas coma is scarce. Thus far, CN and CO are the only species seen 
in the coma, and both were detected only once \citep{bus93,wom99}.  The optical CN detection of $\sim$ 4\ee{25} mol
s$^{-1}$ at 11 AU
correlates to a high production rate for a comet at 11 AU, and $>$4 times greater than Hale-Bopp at 10-12 AU 
\citep{rauer03}. Emission from CN was searched for when Chiron was near perihelion at $\sim$ 8.5 AU, and was 
not detected down to a limit that is $\sim$ 25\% lower than what was observed at 11 AU \citep{rau97}. 
Chiron was reported to be undergoing an outburst during both
observing runs, so it is puzzling that CN was detected only once.
Thus, the CN production in Chiron may be variable and not well-correlated
with either the dust production rate or heliocentric distance. 
The millimeter-wavelength CO detection in Chiron is consistent 
with a production rate of $\sim 10^{28}$ mol s$^{-1}$ at 8.5 AU \citep{wom99}, which is also similar 
to Hale-Bopp at the same distance \citep{biv02}, see Figure \ref{fig:qco}. The observed CO line-width was 
very narrow and similar to what has been observed for 29P and Hale-Bopp. The CO line-width was FWHM = 0.39 km s$^{-1}$; however, 
the line was not resolved, and this should be considered an upper limit to the
true spectral line-width. The narrow line width is indicative of emission 
from a cool gas, T $<$ 110 K, and is also consistent with a 
dusty CO coma, in which dust heats gas in the near-nucleus region, 
resulting in terminal gas velocities of $\sim$ 0.4 km s$^{-1}$ \citep{boi93}. 

An upper limit of Q(CO) $<$ 2$\ee{27}$ mol s$^{-1}$ was obtained 
over five observing runs from Mar  1998 - Jul  2000, when Chiron was 9 - 11 AU
\citep{boc01}. This limit is almost ten times lower than the production rate 
measured several years earlier by Womack \& Stern (1999). Hence, the CO detected in 1995  
may have been the result of an outburst and not caused by any long-lived activity from Chiron, 
a conclusion also reached by \citet{boc01}. 

A detailed study by \citet{fornasier13} determined that a gas production rate of at least 1-10\ee{27} s$^{-1}$ 
were needed to drive dust production rates derived from optical observations of the coma. This 
amount is consistent with the CO production rate measured at 8.5 AU and comparable to upper limits derived from $\sim$ 9=11 AU.  \citet{fornasier13} also raise CO$_2$ as capable of driving the observed dust activity. Unfortunately, there are no detections or
upper limits to CO$_2$ production rates in Chiron. 

The high color temperature of Chiron's nucleus and the CO and dust production rates are 
predicted by two models with a highly porous nucleus containing dust and gas-laden amorphous
water ice with more volatile ices deeply buried \citep{pri95,cap00}.  In these models, 
crystallization of amorphous ice in the nucleus is proposed to occur near the surface, 
which then releases the trapped CO gas. This may be the preferred mechanism, since at the surface temperature of Chiron \citep[120K, see][]{cam94} CO and CO$_2$ ices would not survive shallower than a few skin depths \citep[1 skin depth is $\sim$20 m, assuming a density of 0.7 g/cm$^3$, see][]{pri08}. Thus, such solid ice phases of these volatile species can be ruled out as significant constituents of the top surface layers, and are probably restricted to deeper
sections of the nucleus.

\subsection{Comet 29P/Schwassmann-Wachmann 1}

Comet 29P/Schwassmann-Wachmann 1's coma has been documented for
90 years. The comet appears to be nearly always active with
numerous reports of dramatic outbursts of 4 to 8 
magnitudes over relatively short time periods \citep[e.g., ][]{roe58,roe62,whi80,hug91,
trigo10}. Originally classified as a short-period comet, it
has a low inclination orbit, and its heliocentric distance ranges between
 5.7 $< r < $ 6.3 AU. Its current orbit is nearly 
circular ($e$ = 0.04) and is relatively near 
the water ice sublimation zone. Dynamical studies show that 29P's orbit
has become more circular over time, and it probably only recently
transferred to this
orbit from one much farther from the Sun, possibly the Kuiper Belt. It
is often considered to have a Centaur orbit \citep[e.g., ][]{fer80,ste89,lev94,fer94}. 

A recent study of 29P's nucleus radius derives R$_{29P}$=30.2$^{+3.7}_{-2.9}$ km \citep{scham15}, 
which we use for later calculations in this paper. 
This puts 29P at about the same size as Hale-Bopp and much smaller than Chiron. 
29P's large asymmetrical dust coma often exceeds 400,000 km across, and 
spiral structures often seen in the coma are consistent with one or more
active areas on the nucleus \citep{roe58,whi80,coc82,rea13}. 
Optical imaging techniques implied that the dust coma had a surface brightness 
profile (see Eq. 1) of $s = -1$ to $-1.5$ \citep{mee93,jew93}, 
consistent with a steady-state, uniformly outgassing
coma subject to some solar radiation pressure.  \citet{jew90} estimated 
a dust mass loss rate of 10 kg s$^{-1}$ and dust expansion velocity of 200
m s$^{-1}$ from images. However, this may be an underestimate
of the total mass loss, since the reflected light 
is dominated by light scattered by grains with sizes comparable
to the observed wavelength. As a result, this technique may undercount
larger grains, which probably play an important role in the comet's 
mass loss. One  model of the grain size distributions indicates a much
higher mass loss rate of $\sim$ 600 kg s$^{-1}$ \citep{ful92}. 

CN is often the first emission feature detected in optical spectra of 
comets far from the Sun, 
so it was a surprise when CO$^+$ was the first gaseous species seen in 29P \citep{coch80}.  
CO$^+$ emission in 29P was initially believed to be explained by 
photoionization of CO; however, the lifetime for this process at 29P's distance from
the Sun \citep{lar80,coch91} is too long  to account for the observed column 
densities.\footnote{The lifetime against photoionization
for CO at 29P's distance ranges from 3 - 8 x10$^{7}$ sec, depending
on the solar activity \citep{fox89}. At the average rate of 5\ee{7} sec, 
CO molecules moving at 0.45 km s$^{-1}$ away from the nucleus would 
travel $\sim$ 0.1 AU  from the comet before being photoionized, 
which is well beyond the region where CO$^+$ is observed \citep{fes01}.}
Collisional ionization of CO from 
high energy solar wind electrons has been proposed, instead, to explain formation 
of the CO$^+$ in 29P \citep[see ][]{coch91}. 
CN was eventually detected in 29P in Dec 1989 with a line strength weaker than the
neighboring CO$^+$ lines \citep{coch91,Coo05}. 

The rotational period of 29P's nucleus is unknown, with measurements ranging 
from 14 hours to longer than two months \citep{milesheat16,iva15,Mor09,mee93,sta04,jew90,whi80}. 
The large uncertainty in 29P's rotation period is a significant problem for constraining 
outgassing models. Since 29P's orbit has a very low eccentricity, the surface experiences a 
nearly constant rate of heat forcing (and cooling rate). Thus, the variations in illumination
for any given patch on the surface are predominantly a function of period and obliquity.
The co-latitude and co-longitude dependence drives the sub-surface activity (neglecting 
lateral heat conduction). The uniqueness of 29P's activity pattern may 
be the result of its relatively circular orbit and a non-simple spin state (possibly 
a very slow rotator with a highly oblique spin pole). 

The first direct evidence for a volatile that could generate a coma in 29P
was the detection of CO gas \citep{sen94}. A production rate of Q(CO) = 3.0 x 10$^{28}$
molecules s$^{-1}$ was derived, 
which corresponds to a mass loss rate of $\sim$ 2000 kg s$^{-1}$. Since
this is much greater than the dust mass loss rate of  10 - 600 kg s$^{-1}$, it is assumed 
that CO is driving the observed activity. Upper limits
of Q(CH$_4$) $<$ 1.3x10$^{27}$ mol s$^{-1}$ ($<$ 5\% of CO) were
determined from infrared spectra of 29P \citep{pag13}. 

Similarly, sensitive upper limits to CO$_2$ emission show that 
this species is not dominant in 29P's coma with Q(CO$_2$) $<$ 80 Q(CO) \citep{oot12,rea13,bau15}. 
We conclude that N$_2$, CH$_4$ and CO$_2$ are produced in minor amounts when compared to CO. 
Comet 29P seemingly always has at least a minimal CO production, of
Q(CO) $\sim$ 3\ee{28} mol s$^{-1}$, which is often referred to as the 
quiescent or non-outbursting state \citep{fes01}. 

Sporadic outbursts also occur which increase the comet's CO production by up to a factor 
of seven (see Figure \ref{fig:qco}), even though the comet's orbital distance does not change very much during this time. 
Clearly, CO outgassing is tied to the outbursting mechanism in 29P. Surprisingly, 
no correlations were found between observed CO$^+$ emission and the outbursting state, as evidenced by optical observations 
of the coma. The production
mechanism for CO$^{+}$ in this comet is not clear \citep{coch80,coc82}. Simultaneous observations of CO and 
CO$^+$ are needed to better understand coma chemistry.

The CO activity in 29P was proposed to be released from a relatively
small active area on the surface by \citet{sen94}, either by sublimation of CO, or as trapped in water ice. 
However, detailed models of frozen CO sublimating (whether pure or adsorbed onto water ice) 
does not reproduce the observations well \citep[i.e. ][]{enz97} and the surface is too warm for CO ice to survive
for very long. 
Different models exist that incorporate energy provided by HCN polymerization and the crystallization
of amorphous water ice \citep{gro98} or melting and dissolution of gases \citep{milesheat16}; 
however, they do not include a full account of the energetics, so these models are of limited value. Moreover, the
dust outflow velocities predicted are approximately three times lower than what was measured 
from HST WFPC2 imaging data \citep{fel96}. The model which best explains the 
CO production rates to date is one that includes crystallization of amorphous 
water ice with trapped CO gas and pockets of CO ice, 
a highly tilted rotation axis and surface erosion \citep{enz97}. In this model, CO 
production primarily comes from release as a trapped gas when crystallization occurs 
in amorphous ice in the sub-solar area. This model also explains outbursts with surface 
erosion where dust grains are released from the ice-dust matrix by the escaping gas. 

Further clues about the physical state of the coma are given by the millimeter and infrared spectra of CO. The CO spectral 
line shapes in 29P (in both the quiescent and outbursting states) are frequently characterized by two very narrow 
(FWHM line-width $\sim$ 0.10 to 0.15 km s$^{-1}$) velocity components, which are found at $\sim$ +0.5 km s$^{-1}$ 
and -0.3 km s$^{-1}$ with respect to the
cometocentric velocity \citep[][see Figure \ref{fig:sw1}]{cro95,biv97a,fes01,gun02}.
Two emitting regions have been proposed to explain the double-peaked (red and blue-shifted) 
CO line profile: a day-side and night-side region \citep{cro95}. 
In contrast, the two peaks are interpreted by \citep{cri99} as due to temperature distributions on the surface, and 
not the spatial location of vents. Still another idea is that the line-shape may be partly due to under-sampling 
the CO emitting 
region, which would preferentially detect the gas with the greatest Doppler shifts \citep{fes01}. Finally, the two lines
could be caused by two different sources of CO: a nuclear and an extended source originating from icy grains in the coma  
\citep{gun02}. 

Raster (point-by-point) mapping data with the SEST 15-m telescope
indicated that there was a substantial secondary source of CO in the coma, which 
was hypothesized as being due to the release of CO from icy grains in the 
coma in 29P \citep{fes01,gun02}. In this scenario, the sunward emission from 
the nucleus gives a strong blue-shifted peak, and the isotropic
emission of the extended source gives a so-called ``horned" appearance due to 
under-sampling of the icy grain emission source by the radio beam. The blue-shifted part of the ``horn" is blended with the sunward peak.  With this model, the extended source produces $\sim$ 3 times the 
amount of CO as the nuclear source and is the dominant source of CO emission. 

Subsequent mapping of CO emission with the IRAM 30-m dish did not show evidence for such a widespread secondary source \citep{gun08}. 
However, the data showed an asymmetric localized feature in the coma with 
``excess emission.” This is evident in both the dust and gas comae and extends 0.5 to 2 arcminutes south of the nucleus \citep{scham15,gun08},
which corresponds to a projected distance starting at $\sim 1.1\ee{5}$ km from the nucleus (we find that the
published number in \citet{gun08} is too high by a factor of ten). This is sometimes also referred to as a jet feature. 
Thus, it is possible that 29P split off a smaller piece or fragment during 2003 - 2004. Interestingly, CO emission was detected from both the main and fragment sources by \citet{gun08}. The fragment was found to have a slightly smaller blueshift (-0.18 km/s) than the main nucleus component (-0.35 km/s). The spectral line profile of the secondary also showed the classic horned appearance, consistent with the spectra obtained at the primary position. Follow-up high angular resolution spectral mapping, or interferometry, would be invaluable for studying this ``excess emission" phenomenon in 29P.

The true nature of this enhanced CO emission region in the coma is not known. A long lasting, but largely hidden, fragment or secondary component could be an important additional source of CO in 29P's coma \citep{gun08}. A secondary component would also complicate efforts to measure the rotation period of the main nucleus, which has so far remained undetermined. 

We note that 174P/Echeclus, another Centaur that is sometimes active, reportedly had a distinct secondary source, which may have recently split from the primary nucleus \citep{rous08, fern09}. This secondary extended emission in Echeclus was at a similar distance from the nucleus: the angular separation is a factor of 
$\sim$ 3 times, and the derived spatial separation is $\sim$ 1.5 times that of 29P's excess emission \citep[calculated from][]{rous08,gun08} using the correct distance (see previous paragraph). Processes acting 
at 29P may be similar to those at Chiron, the other Centaur in this study, producing secondaries and rings. 

Infrared and millimeter wavelength spectra also showed that the CO inner coma gas in 29P is at an extraordinarily low rotational temperature of $\sim$ 5K \citep{gun08,pag13}, making it one of the coldest gases known in the the solar system. As discussed in \citet{gun02,pag13}, if CO molecules start out from the nucleus with T$_{rot} \sim$ 5K, they will change excitation conditions over time and reach fluorescence equilibrium within $\sim$ 2 days. 
Therefore, many millimeter wavelength observations with larger beamwidths are recording spectra from molecules which are close to fluorescence equilibrium with a higher rotational temperature of 10-15K.

29P's coma is predicted to have a significant amount of
water due to the sublimation of icy grains, if water ice is present at the surface
and if grains are ejected \citep{cri99}. Initially, searches showed no water down to a limit
of Q(OH) $<$ 2-3\ee{28} mol s$^{-1}$ \citep{coch91,fel96}. 29P was observed with the Herschel Space Observatory, and preliminary results claim a detection of water and HCN with line profiles that are consistent with being emitted from icy grains in the coma, but no values have been published yet for the production rates \citep{boc10,boc14}.

\section{Discussion and Comparison of Distantly Active Comets}

In this section, we review the similarities and differences
in the observational data of Hale-Bopp, 29P and Chiron, and discuss the models that 
best fit these data. We focus on the spectroscopic results, mainly production rates and line profile characteristics. 
Representative production rates of important species in 29P, Chiron and Hale-Bopp are summarized in 
Table \ref{tab:volatile}. 
In particular, we 1) examine the production rate ratios of $CN/CO$ (and $HCN/CO$) and 2) $CO_2/CO$, 3) discuss clues from 
spectral line profiles of volatile species, and 4) recalculate CO production rates for all 3 comets with 
consistent modeling assumptions. We then 5) summarize observational constraints to models of distant activity and 
6) analyze Q(CO) for all three objects when adjusted for nucleus size and heliocentric distance. 

\subsection{Production Rate Ratios: CN/CO and HCN/CO}

CN and CO are the only gaseous species observed in all three comets, and we briefly compare
their relative production rates to test models of CN and HCN in comae and to assess whether HCN is a likely 
competitor for triggering activity in these objects. Using values from \citet{jew96,fitz96,rau97,biv02}, 
we calculate that the ratio Q(CN)/Q(CO) was $\sim$ 0.003 in Hale-Bopp at $\sim$ 6.8 AU pre-perihelion, and the same value at 4.6 AU 
post-perihelion (see Table \ref{tab:cnco}). We derive the ratio Q(HCN)/Q(CO) $\sim$ 0.004 at 6 AU post-perihelion using values from \citet{rau97,biv02}. Interestingly, Q(HCN)/Q(CO) $\sim$ Q(CN)/Q(CO), even at 
somewhat different heliocentric distances. This is consistent with work by \citet{rauer03,fray05,boc08}, which conclude 
that Hale-Bopp's
CN largely originated from HCN photolysis. Additionally, it is clear that HCN was produced in much smaller amounts than CO in Hale-Bopp at large heliocentric distances, and thus does not compete with CO in triggering activity. 

Unfortunately, it is not clear what the CN/CO production rate ratio is for Chiron, because CN and CO were not measured at the same time. However, if we use  the CN value at 11 AU and CO measurement at 8.5 AU, we get a similarly low value of Q(CN)/Q(CO) $\sim$ 0.003. However, we stress that these were not simultaneous measurements for Chiron, and hence the ratio is probably of limited value. 

CN is only rarely observed in 29P and, when seen, it appears depleted with respect to CO, when compared to Hale-Bopp, which we use as a proxy for an unprocessed new comet. Using the CN production rate from \citet{coch91} and the average non-outbursting value of Q(CO) $\sim$ 3\ee{28} for 29P, we calculate a production rate ratio Q(CN)/Q(CO) $\sim$ 0.0003, which is almost ten times lower than what is derived for Hale-Bopp at 4 and 6 AU. By this metric, 29P appears to have much less CN (and presumably HCN) than CO when compared to Hale-Bopp and possibly also Chiron. As mentioned previously, 29P was observed with the Herschel Space Observatory, and preliminary results claim detections of H$_2$O and HCN \citep{boc10,boc14}. Once HCN production rates are published from these Herschel data, they will be useful for comparing with measured CO and CN production rates to test models of release of all three species from the nucleus \citep{mckay12}. It will be also interesting to see whether the Herschel data support CN arising from HCN in 29P. The relative amounts of HCN and CO produced from 29P could provide clues to the comet's formation environment. For example, if HCN and CO are both incorporated into amorphous ice, and if the cometary material was predominantly from regions where CO was more likely to be adsorbed onto forming grains of amorphous water ice, then HCN depletion would indicate Kuiper Belt formation distances, rather than closer-in Saturn distances.

\subsection{Production Rate Ratios: CO$_2$/CO}

Another important marker in comae is the relative amount of CO$_2$ and CO, 
which is a useful test of thermal evolution and/or formation models 
\citep{feldman97,belton09,ahearn12}.
Although recent surveys show that CO$_2$ $>>$ CO for most comets \citep{oot12,reach13,bauer15}, CO drives the activity
for Hale-Bopp (beyond 4 AU) and 29P. Indeed, these two comets appear to be ``CO-rich" when compared to most other comets. 

CO$_2$ was directly detected in Hale-Bopp's coma at 4.3 microns using the Infrared Space Observatory at 4.6 AU pre-perihelion and 4.9 AU post-perihelion \citep{cro99}. CO$_2$ was also indirectly detected in Hale-Bopp with similar values via CO Cameron band emission. These CO bands predominantly arise from CO$_2$ dissociation \citep{weaver97} but also have contributions from electron impact excitation of CO \citep{bhardwaj11}. CO$_2$ production rates can be measured via high spectral resolution ground-based observations of O[I] when cometary and telluric emission is significantly separated by Doppler shift \citep{mckay12,raghuram14,decock15}, but this technique was not used on Chiron, Hale-Bopp or 29P. Using directly measured CO and CO$_2$ production rates at approximately the same time from the literature, we calculate production rate ratios of Q(CO$_2$)/Q(CO) $\sim$ 0.1-0.3 for Hale-Bopp between 4-5 AU (Table \ref{tab:co2co}). Searches for CO$_2$ in 29P yielded upper limits of Q(CO$_2$)/Q(CO) $\sim$ 0.01 \citep{oot12,rea13}, roughly ten times lower than for Hale-Bopp at $\sim$ 6 AU. Clearly, at these large heliocentric distances, CO$_2$ is not a significant driver for activity in either Hale-Bopp or 29P.
Interestingly, the Hale-Bopp and 29P Q(CO$_2$)/Q(CO) ratios are significantly lower than what is reported for many other periodic or dynamically new comets \citet{feld97,ahearn12,bauer15,reach13,oot12}. It is important to note that most of these measurements of other comets were obtained at heliocentric distances of 1 to 2.5 AU, which is much closer to the Sun than the CO$_2$ measurements in 29P and Hale-Bopp. It is possible that the production rate ratios of CO$_2$ and CO vary in distance from the Sun for an individual comet. The production rate ratios may not reflect the nucleus abundances, but rather the activity mechanism dominant at the time of observations. For example, at 29P's distance from the Sun, the effective surface temperature (black-body) is roughly 120 K, which straddles the threshold for water ice amorphous-crystalline transition. Due to differences in volatility (sublimation pressures and rates), released CO will immediately flow outside, while released CO$_2$ would diffuse in the nucleus and may even re-condense at various depths. The presence of ice and clathrate reservoirs, for CO$_2$ and CO, could introduce alternative pathways. However, the location of these components will have to be very close to the surface, within $\sim$1 skin depth, in order to match the quiescent activity of 29P.       

Indeed, the measured Q(CO$_2$)/Q(CO) ratio increased as Hale-Bopp approached the Sun, which may be partly due to higher sublimation temperature for CO$_2$ 
(see Table \ref{tab:svolatile}) being reached by the comet. As part of a large infrared survey of comets that included 9 comets beyond 4 AU, \citet{kel13}  
calculated each comet's ef$\rho$, which is a quantity used to approximate the dust production rate. The study showed that log 
Q(CO$_2$)/$\epsilon$f$\rho$  broke into two groups of behavior for comets inside and outside 4 AU (see Figure 9 in \citet{bauer15}). The authors of that survey suggest that within 4 AU, CO$_2$ activity may be endogenic with the dust. This may be attributable to different source regions (surface versus sub-surface) for cometary CO$_2$ and CO emissions. Further searches for CO and CO$_2$ emission from other distantly active comets are needed to test models of volatile production. Hale-Bopp and 29P appear to be CO-rich comets. 

\bigskip

\subsection{Spectral Line Profiles}

Additional information is contained in spectral line profiles about how gas is emitted from a comet nucleus. 
For example, consider that the millimeter wavelength spectral line profiles of CO for Hale-Bopp (when beyond 4 AU), 29P 
and Chiron are all very narrow ($<$ 0.5 km/s), and sometimes are blue shifted by a small amount, $\sim$ 0.3-0.5 km s$^{-1}$.
The numerical values of the velocity shifts are consistent with CO escaping from the nucleus at velocities corresponding to hydrodynamical 
escape for $v \sim$ 0.4 km s$^{-1}$ \citep{fes01,gun03}. The spectra's blue-shifts indicate that the CO molecules' velocity 
vector is primarily toward the Sun (and hence, the Earth, due to typically small phase angles),
and the emission can be understood if it occurs on the sunward side of the nucleus. Such narrow lines indicate a narrow distribution of fluxes coming off the nucleus, which in turn probably indicates a singular source for the CO, at least in terms of the temperature and depth. There is a narrow range of thermal velocities that the CO molecules experience, once released, and there is a specific length scale that these molecules traverse before leaving the surface.  

Sublimation from an icy grain halo is expected when an active comet
is $\sim$ 4-6 AU from the Sun. This occurred in Hale-Bopp's coma 
at $\sim$ 5 AU from the Sun, and there is strong evidence that this 
develops in 29P and around other comets, such as Bowell, at 5 AU.
This is the likely origin of most of the CH$_3$OH, OH, HCN, H$_2$CO and H$_2$S 
emission in Hale-Bopp beyond 4 AU, as is illustrated in Figure ~\ref{fig:cometh}. 
Interestingly, as Hale-Bopp moved closer to the Sun (from 5 to 4 AU) 
the linewidths of methanol increased from $\sim$ 1 to
2 km s$^{-1}$, while the CO linewidths remained relatively constant, which 
further illustrates the differing behavior for CO and CH$_3$OH at large heliocentric distances \citep{wom97}.

\subsection{A Consistent Set of CO Production Rates}

\begin{figure}
\epsscale{.9}
\plotone{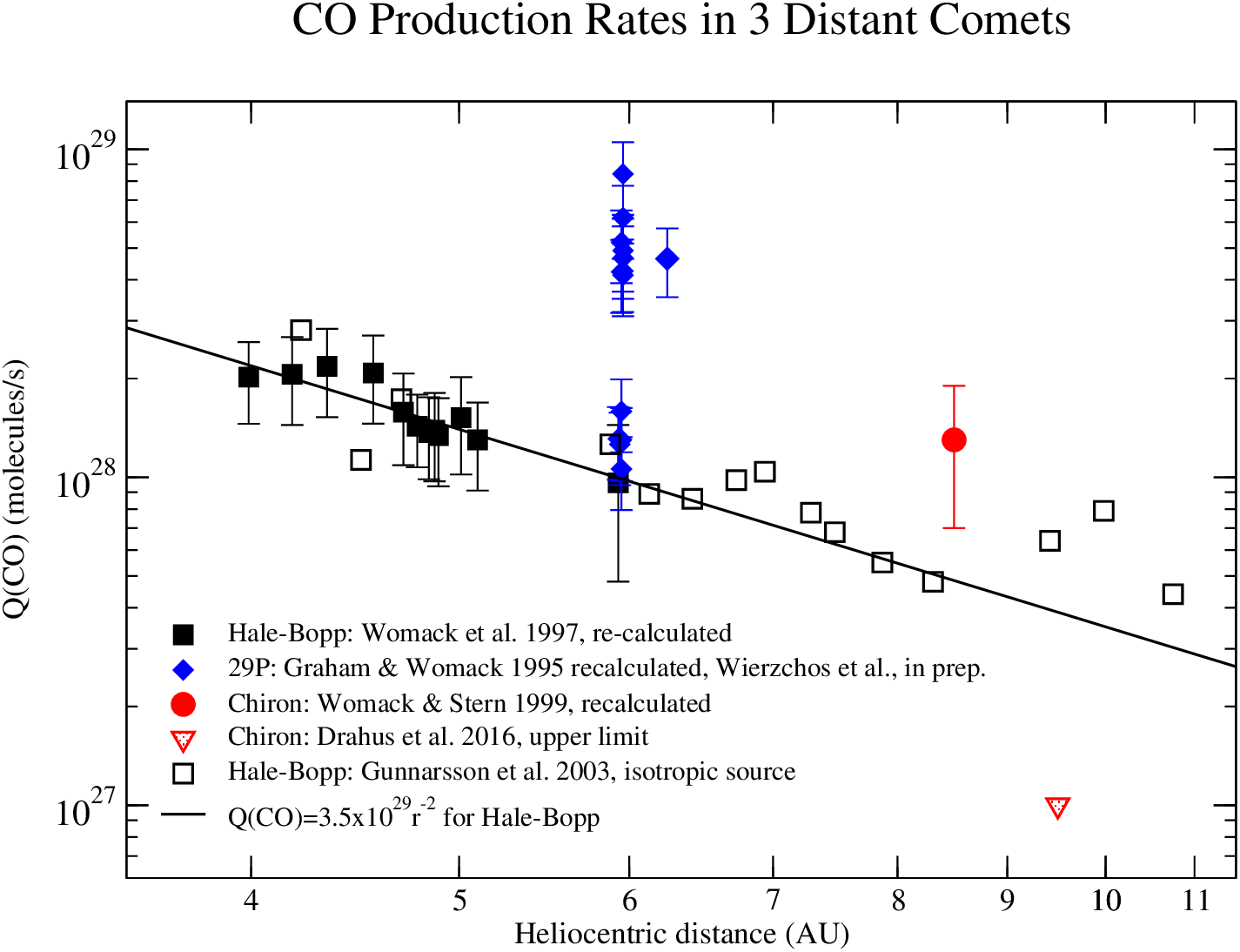}

\caption{
\label{fig:qco}
CO production rates from three distant comets are plotted against 
heliocentric distance. The observations of Hale-Bopp and 29P are consistent with a nominal production rate of Q(CO)=3.5\ee{29}$r^{-2}$ superimposed with sporadic outbursts.
}
\end{figure}

\begin{figure}

\epsscale{.9}

\plotone{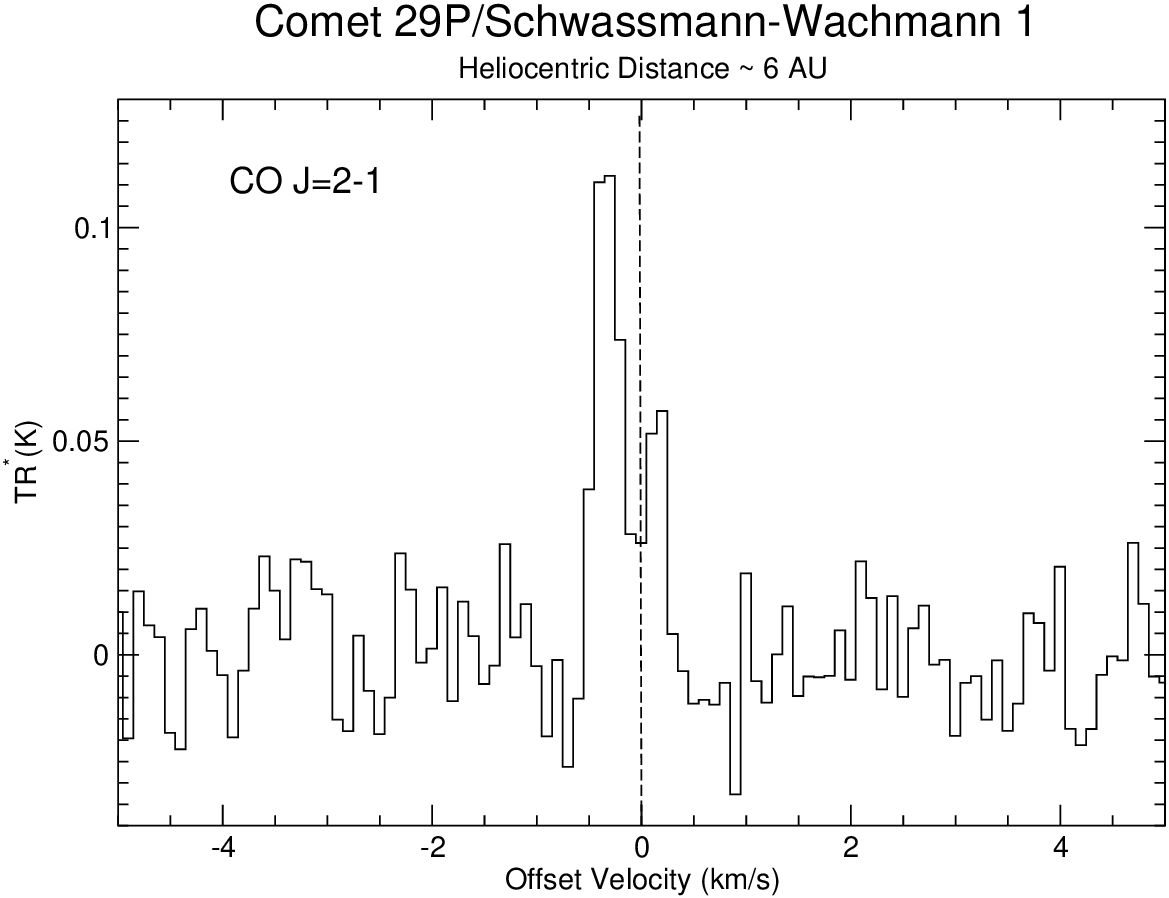}
\caption{
\label{fig:sw1}
Spectrum of the CO J=2-1 rotational line in comet
29P/Schwassmann-Wachmann 1 obtained from 1995 Dec 9--11 UT with the NRAO 
12-m telescope \citep{gra95}. The spectral resolution was 100 kHz, which corresponds to 
0.13 km/s per channel. The dotted line indicates the comet's
rest frame velocity. Two narrow velocity components are seen in this 
spectrum and in many other observations of this comet.
}
\end{figure}

Production rates for CO in these objects were published previously by several groups over 20 years with different models and assumptions, including \citet{sen94,cro95,wom97,wom99,biv02,gun03,pag13,dis99}. Differences in 
observing and modeling techniques make it difficult to assess similarities and differences of the 
production rates among the comets. In order to better compare the CO production, and probe the importance of nucleus size and thermal history, we derived the CO production rates using our own data for Hale-Bopp, 29P and Chiron in a consistent manner. The data for Hale-Bopp and Chiron were previously published using simpler models \citep{wom97,wom99}. Most of the data for 29P are new and were obtained during 2016 Feb to May using the Arizona Radio Observatory 10-m Submillimeter Telescope (SMT)(Wierzchos et al., in preparation).  The 29P CO data are part of
a separate project and a full analysis will be published later. For now, we present the dates, heliocentric distances and derived 
production rates in Table \ref{tab:29pqco}. One additional data point for 29P was taken from \citet{gra95}. We assumed isotropic outgassing of CO from the nucleus, rotational and excitation temperatures of \citet{biv02} and expansion velocities from \citet{gun08}. The excitation model includes both collisional and fluorescence excitation in a manner similar to Crovisier \& Le Bourlot 1983 and 
\citet{biv97}. Applying a cone ejection model would reduce the numbers in Figure \ref{fig:qco} by ∼ 40\%, but would not otherwise change the results. In Figure \ref{fig:qco}, we plot the new CO production rates for Hale-Bopp, 29P and Chiron. The Hale-Bopp and 29P production rates are comparable to values derived by \citet{gun03,gun08}. In Figure \ref{fig:qco} we also include Q(CO) values for Hale-Bopp taken from \citet{gun03}, to extend the heliocentric range for Hale-Bopp; these values were not recalculated, because they used the same modeling parameters as our new model. The upper limit for Chiron was taken from \citet{drahus16}.

The Hale-Bopp CO production rates in Figure \ref{fig:qco} are all pre-perihelion and follow the relationship Q(CO) = 3.5\ee{29}$r^{-2.2}$ molecules/second, where $r$ is heliocentric distance in astronomical units (AU). This
heliocentric dependence agrees with what was found by \citep{biv02} for pre-perihelion data of Hale-Bopp. Later post-perihelion observations of the comet out to $\sim$ 11 AU showed that Q(CO) followed a slightly less steep $r^{-2}$ dependence \citep{biv02,gun03}. The pre-perihelion slope is steeper than one would expect for a mechanism driven solely by insolation, which would produce a $r^{-2}$ dependence. The steeper slope may indicate that there is an additional source of energy in this regime, such as crystallization of water ice. Based on the discussion and analysis of \citet{gun03}, we extrapolated the Q(CO) relationship for Hale-Bopp out to 11 AU, assuming an $r^{-2}$ dependence to indicate the likely CO production rates at larger heliocentric distances.

When adjusted for heliocentric distance, CO production rates in three distantly active comets are remarkably
uniform, despite different thermal processing histories (long-period vs. Centaurs)
and different nucleus sizes. We explore the apparently similar CO production rates in the next section.

\subsection{Observational Constraints to Existing Models of Distant Activity}

The models that best explain the quiescent activity of Hale-Bopp and 29P 
are similar. A porous comet nucleus comprising a mixture of ices and grains with a 
dust mantle at the surface best fits the data obtained so far \citep[e.g., ][]{enz97,enz99,cap02}.
In these models, crystallization of amorphous water ice near the surface 
and subsequent release of trapped CO molecules is the
dominant cause of activity for 29P and Hale-Bopp within 7 AU. 
Activity in Hale-Bopp beyond ∼ 8 AU is more likely caused by sublimation of near-surface CO ice, since the 
heating rate is too low for crystallization to proceed in any appreciable rate.
The survival of near-surface CO ice may be aided by the low thermal conductivity maintained in very porous surface layers. 
Existing in-situ measurements of comet nuclei point to a highly-porous bulk, as for the Rosetta and Deep Impact measurements of 
comets 67P and 9P, respectively (Sierks et al. 2015, A'Hearn et al. 2005). However, a porosity gradient has been inferred 
for 67P (from the CONSERT sounding experiment), with an upper limit on near-surface porosity of 30
In any case there should exist a long-term reservoir for highly-volatile species, such as CO. The distinction 
between the effects of an amorphous/crystalline water ice phase and a gradient in layering and insulation falls 
outside the scope of this work and will be pursued by systematic modeling efforts.     

Moreover, the
CO emission in 29P and Hale-Bopp has sunward velocity shifts at large heliocentric
distances. This implies that
the CO released by both methods (sublimation of CO ice and release of trapped CO during the
crystallization of water ice) occurs from the sub-solar regions of the nuclei. 
For Chiron, the surface temperatures are too warm for solid CO or CO$_2$ to be
near the surface and sublimate, so instead, crystallization of water ice and 
release of trapped gases is the more likely model \citep{pri95,cap00}.
A high obliquity for the rotation axis would allow Chiron's sub-solar regions to be heated
enough by the Sun to reach the measured color temperatures, which is also high enough
for crystallization to occur. These models also
reproduce the quiescent dust production rates of the comets. 

However, we note that recent spacecraft data have revealed how geologically diverse comets are. This surface geology and morphology evolution and the connection to the sub-surface layers are a function of both origin and evolution. Thus, it is not just the specific diffusivity and conductivity of the icy and silicate components, but also the configurations, mixing ratios and dynamics of the different constituents that become important. Since comets are significantly porous \citep[e.g., ][]{patz16,AHearn11,AHearn05}, the way the porous medium evolves together with compositional and structural changes (phase transitions, gas flow, cavities, small impacts, out-gassing) affects the determination of bulk transport properties. Models of these interactions are inherently susceptible to some degeneracy in the choice of parameters. The more chemical and physical information we have about the nucleus and the behavior of the coma at various distances, the better we can constrain the modeled profiles and compositions.

Gaseous outbursts can be explained if a region unusually rich in CO 
ice sublimates when exposed to the Sun, or if a crystallization burst of water ice occurs. 
Either process will cause an increase in CO and dust
production (if there is much dust in that region). 
Therefore, large dust jets emerge from cracks in the surface after being triggered by a gaseous
outburst, while the bulk of the observed CO (that generates the quiescent activity) 
percolates through the comet surface that is at or near the sub-solar spot. Outbursts in
29P, in particular, are possibly triggered by surface erosion of the dust mantle \citep{enz97}.
For Chiron, sublimation of CO ice pockets is
not likely, since, as discussed previously, the surface temperatures are too warm for 
CO ice to be present. Instead, amorphous water ice crystallization
bursts are considered to be the most likely cause of outbursts for Chiron.

The mechanisms proposed to explain the distant activity in 
Hale-Bopp, 29P and Chiron are possible in any comet which contains
amorphous ice with trapped CO gas, as well as frozen CO in their nuclei, as long as
these materials are sufficiently
near the surface to be heated to the crystallization and/or sublimation temperatures.
One might wonder why most comets are not reported to be active until within 3 AU of the Sun. One 
possibility is that most short-period comets exhausted their reservoirs of CO (and other supervolatile ice) and converted 
their amorphous ice to the crystalline phase many millions (or even 
billions) of years ago. Any remaining frozen CO and amorphous ice is
too deeply buried to be heated above the crystallization and sublimation temperatures
during their orbits. Another important point is that the nuclei of Hale-Bopp, 29P and 
Chiron are larger than those of most typical comets, and this is probably responsible for
much of the distant activity. The larger volume exposes more material to heating, 
which provides more opportunities for sublimation and crystallization. 
Some smaller, long-period comets may exhibit CO production beyond 3 AU, but the quantities
are too low to be detected with current equipment. 

Orbital history is thought to play a role in devolatilizing comet nuclei. Hale-Bopp's orbit places it in the 
long-period comet category, which means it has experienced few heating cycles over its lifetime, especially when compared
to Chiron and 29P. 
It is dynamically young and has probably experienced very different perihelion/aphelion evolution than low-inclination bodies in the outer solar system (Bailey et al. 1996). As such, the reservoirs of volatile species in the nucleus have been processed in a different manner and could have remained more accessible to subsequent apparition.    Large gas fluxes from previously untapped reservoirs of volatile species, which can efficiently entrain and drag dust grains from various depths in the nucleus, could have been the main source of the distinctive brightness of Hale-Bopp. The highly-volatile nature of the gas (and perhaps larger abundances of CO) may have been the main cause for very distant activity, both pre- and post-perihelion. 29P orbits much closer to the Sun, at $r \sim$ 6 AU, and thermal evolution from previous orbits probably depleted any highly-volatile solid ice from at least the first 100 or so meters. Accessible solid ice layers are probably rare, and thus not the main driver of activity for 29P. This is also probably true, to some extent, for Chiron, which orbits between 8.5 - 13 AU. By this measure alone, Hale-Bopp should have produced significantly more CO than 29P or Chiron. However, as Figure \ref{fig:qco} shows, the CO production rates are consistent between the comets. 29P produces CO at approximately the same rate as was observed in Hale-Bopp when it was at $r \sim$ 6 AU. Hale-Bopp's and 29P's very different orbital histories appear not to be reflected in their CO production rates at 6 AU. Chiron is a little more complicated to sort out. The CO emission detection in Chiron at 8.5 AU is consistent with the CO production rate in Hale-Bopp at about the same heliocentric distance post-perihelion. However, it is important to keep in mind that other searches for CO in Chiron led to upper limits which were significantly lower than the detection, which presumably occurred during an  outburst. The non-detection limit made over two years implies that if Chiron does continually emit CO, it is likely at a much lower rate than Hale-Bopp or 29P. Dust production rates estimated from optical-infrared images are also consistent with significantly lower activity overall compared to Hale-Bopp at this distance \citep{rom03,wei03}. Given
the low dust production rates, 
Chiron is not likely to be outgassing significant amounts of another volatile, such as CO$_2$. The Chiron data are consistent with very little constant volatile outgassing punctuated by sporadic outbursts. 

Although Hale-Bopp's long orbital period means that it is substantially less heated/processed than short-period 
comets, it has endured several passages near 
the Sun, which may be evident in comparisons with a true dynamically new comet, making its first perihelion approach. 
Detailed comparisons of observations of Hale-Bopp with such dynamically new comets at large distances would be 
very useful and should be encouraged. Currently, very few direct 
comparisons are possible beyond $\sim$ 2 AU, but one recently studied dynamically new comet, C/2009 P1 Garradd, showed a striking 
difference. Garradd did not show the strong
correlation between CO activity and heliocentric distance as Hale-Bopp. 
Rather than show a symmetric increase and decrease of CO around perihelion, Garradd's CO production 
continued to increase throughout the entire time period that could be studied, out to $\sim$ 2-3 AU post-perihelion \citep{feaga14,mckay15}. Unfortunately, observations of CO emission beyond 4 AU do not exist for comet Garradd. It would be
particularly valuable to compare CO behavior for these two classes of comets when they are beyond the water-ice sublimation limit. 
Also, closer to the Sun, CO may be produced in significant amounts as daughter products from parent species. Such competing sources
for CO production near the Sun complicates efforts to account for how much CO is natal and driving the long-term activity.

Of the three objects discussed in this paper, two that have very different orbital histories (29P and Hale-Bopp) and thus should have very different CO production rates, in fact have very similar CO production rates. In addition, two which are classified as Centaurs with relatively close-in orbits (29P and Chiron) are expected to produce similar amounts of CO had they started with similar compositions. However, they have vastly different CO production rates. This is a very small sample set, obviously, but from studying these three objects we conclude that orbital history alone cannot explain the CO production rates in these objects. Further constraining measurements of CO emission from other Centaurs are needed to explore the role CO outgassing plays in Centaurs.

\subsection{CO Output Adjusted For Nucleus Size}

A comet's size is also expected to be correlated with how productive it is when it is heated by the Sun. 
For larger objects, once amorphous ice starts transitioning, triggered by 
small increases in temperature (from insolation, impacts, chemical reactions, etc.), it will
depend only on the heat wave coming in from the surface. This heat wave is just a 
function of heliocentric distance, if we assume all other properties are similar
(material, porosity, distribution, etc.). 

Larger comets are more susceptible to radiogenic heating, which may lead to devolatilization of the original ice. As discussed earlier, comets smaller than 10 km probably have their initial temperature preserved deeper than about 100 meters. However, 29P, Chiron and Hale-Bopp all have radii larger than 10 km, and so probably experienced some amount of radiogenic heating. Chiron is much larger than 29P and Hale-Bopp and so probably underwent even more radiogenic heating (unless 29P and Hale-Bopp are fragments of larger objects). Consequently, on the basis of size, one might expect Chiron to be more processed, and thus have less outgassing than 29P or Hale-Bopp. On the other hand, if one assumes comet nuclei to be roughly heterogeneous, then bigger comets should have higher outputs when heated. 

CO emission was searched for, and not found, 
in a collection of KBO's and Centaurs by \citet{boc01,jew08,drahus16}. Many of the Centaurs were first observed by \citet{boc01} and
\citet{jew08} and then re-analyzed by \citet{drahus16}, who also made new observations. The authors of the
first survey proposed that if the objects' CO activity followed
the trend of Hale-Bopp's gas activity with heliocentric distance, and is proportional to the
object's diameter, $D$, then the relationship Q(CO) $\sim r^{-2}\times D^2$ should
hold, and CO should have been detected in these very distant objects. The upper 
limits they obtained are significantly below what would
be predicted from this simple scaling law, which they conclude is evidence that 
that these objects underwent significant CO devolatilization since their formation \citep{desanct01}. The authors of the second survey drew the same conclusions about Centaurs likely being depleted in CO using similar arguments, whereas the third survey raised the question of 
whether CO outgassing should be correlated with heliocentric distance or size, but instead a small heliocentric distance might be more important. 29P is a notable exception to the list of apparently CO-depleted Centaurs, and its nearly circular orbit has a relatively small heliocentric distance of $\sim$ 6 AU.

\begin{figure}
\epsscale{.9}

\plotone{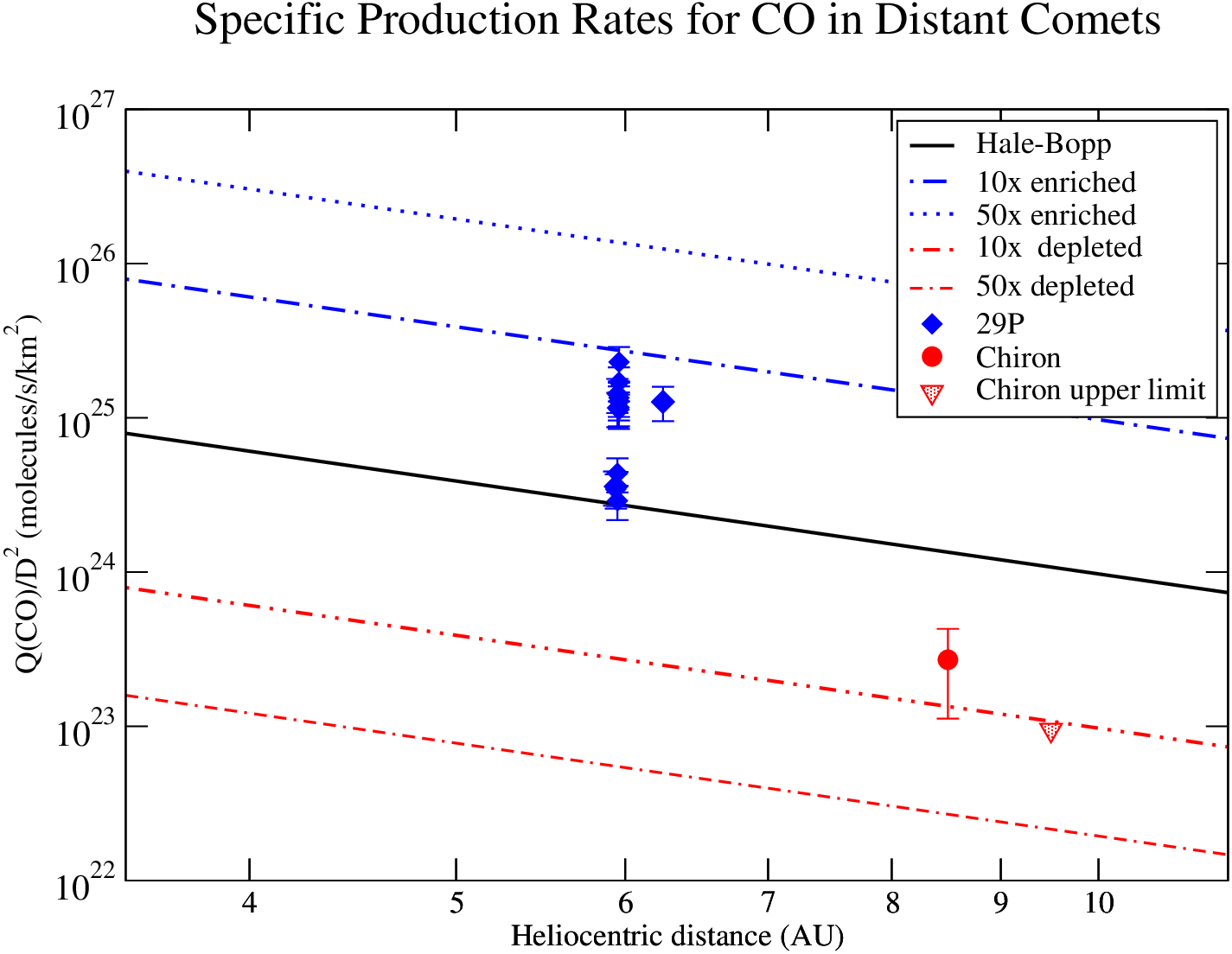}
\caption{
\label{fig:specific}
Specific production rates, Q(CO)/D$^2$, for distant comets. The solid line is 
the specific production rate derived for Hale-Bopp, assuming Q(CO)=3.5\ee{29}$r^{-2}$, and diameter 
D$_{Hale-Bopp}$=60km. Values for Chiron and 29P are plotted assuming production rates from 
Figure \ref{fig:qco} and D$_{Chiron}$=218km and D$_{29P}$=60 km. This plot
shows that, when adjusted for surface area and heliocentric distance, 
29P (in non-outbursting mode) produces CO at approximately the same rate as, and up to ten times larger than, Hale-Bopp.
In contrast, Chiron produces at least 5-15 times less CO than Hale-Bopp. 
This may indicate that 29P entered its current orbit more recently than many models predict, and 
that Chiron is significantly depleted in CO over its initial chemical composition.
}
\end{figure}

To explore this further, we calculated CO ``specific production rates," which are the production rates divided by the square of the nucleus diameter. Specific production rates were calculated for Hale-Bopp, 29P and Chiron using CO production rates from Figure \ref{fig:qco}. Note that the Chiron upper limit at $\sim$ 9.5 AU is 
from \citet{drahus16}, which they recalculated using \citet{boc01} data. The specific production rates are plotted in Figure \ref{fig:specific}. Accordingly, Chiron is approximately 5-15 times depleted in CO when compared to Hale-Bopp, a proxy for a relatively unprocessed 
comet.\footnote{A similar calculation was performed by \citet{boc01} and found Chiron to be depleted by a factor of $\sim$ 35 with respect to Hale-Bopp. This difference in depletion values can be largely explained by different values used for Hale-Bopp's nucleus diameter, D, and CO production rate at 8-9 AU. For example, \citet{boc01} assumed D=40 km diameter (we use 60 km) and they used an older CO production rate value for Hale-Bopp that was later updated by \citet{gun03}, which we also use.} We do not have our own Hale-Bopp data at 8-10 AU, and hence we extrapolated Q(CO) from the $r^{-2}$ fit to the Q(CO) data, which is also consistent with Q(CO) values from \citet{gun03}. We conclude that there is measurably less CO produced per surface area from Chiron than either Hale-Bopp or 29P. This is consistent with radiogenic heating devolatilization being an important parameter in nucleus modeling, since Chiron is much larger than either of the other two comets.


As discussed in the previous section, Hale-Bopp and 29P have very similar CO production rates, even
when corrected for nucleus size. We see no evidence for devolatilization for 29P when 
compared to Hale-Bopp; in fact, when outbursting, 29P outproduces Hale-Bopp by almost a factor of ten. 
This may point to 29P being a more recent entrant to the inner solar system than Chiron.

Since Chiron does not get closer than $\sim$ 8.5 AU, where T$_{bb} \sim$ 100K, it can maintain a slow 
crystallization rate of $\sim$ 1000 years (and rate of trapped volatile release),
if we assume maximum efficiency.  Perhaps Chiron never experienced large temperature
excursions to its surface, since the perihelion was never much inside of its current one. 
However, by considering the noted higher color temperature and low CO production rate, we find it consistent with Chiron's being appreciably devolatilized over a fresher object like Hale-Bopp. Also, Chiron's interior may be warmer due to early radiogenic heating and slow cooling rates, as it is a larger object than both Hale-Bopp and 29P.





\section{Conclusions}

All Hale-Bopp, 29P and Chiron have in common that:  1) They exhibit long-term 
quiescent activity beyond 4 AU that is punctuated by sporadic outbursts; 2) 
Emission from CN and CO was seen in their coma; 3) The CO line profiles were 
narrow, $\Delta$v(FWHM) $\sim$ 0.3-0.5 km s$^{-1}$; 4) They have spent
most of their lifetime far from the 
Sun;\footnote{Chiron and 29P have only recently entered their
current orbits and may be precursors to Jupiter Family comets \citep{fer94}, 
and Hale-Bopp has an aphelion of $\sim$ 360 AU, spends most of its time
at very large heliocentric distances.} 5) They have larger-than-usual nuclei.

There are notable differences among the comets, including: 1) The CO emission had blue-shifted velocity components of $\sim$ -0.4 km/s (sunward) for 29P and Hale-Bopp, but almost no shift for Chiron's emission; 2) CO was seen only once in Chiron, but was repeatedly observable for long-term study in Hale-Bopp and 29P (and 29P is observable, even now); 3) Hale-Bopp had much higher dust production rates than seen in 29P or Chiron.

Sublimation or escape of molecules from a grain halo at $\sim$ 5-6 AU appears to be a typical development in distantly active comets, and may be the main source of emission of OH, HCN, CH$_3$OH, H$_2$CO, and H$_2$S in Hale-Bopp between 3 -- 6 AU.
This may also explain the origin of OH and CN molecules in other distant comets.  Spectral line
profiles and spatial mapping should be used whenever possible to determine a nuclear or extended
origin for volatiles in distant comets. Observed radio line profiles are consistent with the development and sublimation of icy grains in the coma at about 5-6 AU, and this is probably a common feature in distantly active comets, and an important source of other volatiles within 5 AU. Note that narrow CO line profiles for all three main bodies in this study indicate a nuclear origin for 
CO beyond $\sim$ 4 AU, and not an extended source, as with these other molecular cases.

Our calculated CN/CO and CO$_2$/CO production rate ratios for Hale-Bopp and 29P show that HCN and CO$_2$ were both produced in significantly smaller amounts than CO, and can only be minor contributors to distant activity of these two objects. 
For Chiron, no limits exist for CO$_2$ emission, and CN and CO were both detected only once at the $\sim$ 3.5$\sigma$ level. The CN measurements are consistent with HCN being an insignificant contributor to Chiron's coma. The CO data correspond to a gas production rate 
high enough to drive the observed dust coma activity; however, the CO detected may have also resulted from an outburst and we 
cannot rule out significant contributions from undetected CO$_2$ or other supervolatile gases.
We point out that the CN/CO production rate ratios that we derive from the literature 
are remarkably similar in all three distantly active comets, which warrants follow-up study.

When adjusted for heliocentric distance, our analysis shows that 
CO production rates in three distantly active comets are remarkably uniform, despite different 
thermal processing histories (long-period vs. Centaurs)
and different nucleus sizes. Specific CO production rates, Q(CO)/D$^2$, were calculated in order to take into account nucleus size, which is predicted to play a large role in cometary activity. We find that orbital history does not appear to play a significant role in explaining the CO production rates for 29P and Hale-Bopp, which
confirms the conclusion in \citet{ahearn12} for comets within 2 AU. 29P outproduces Hale-Bopp at the same heliocentric distance, although it has been subjected to solar heating far longer, and models predict it should have been devolatilized over a fresher comet, like Hale-Bopp.  However, we may see evidence for the relevance of nucleus size when considering specific production rates. If CO is present in Chiron's coma on a long-term basis, it must be in relatively small amounts and consistent with a CO depletion by at least 5-15 times over Hale-Bopp, perhaps due to increased radiogenic heating made possible by its much larger size or its higher processing due to orbital history. 

The model which best reproduces the distant activity of 29P and Hale-Bopp is a 
comet nucleus composed of dust, amorphous water ice and smaller amounts of trapped CO
and CO$_2$ trapped in the amorphous ice and a build-up of a thin porous dust mantle 
\citep{pri97apj, pri97}. Within $\sim$ 4 - 7 AU, the water ice undergoes 
crystallization near the surface and releases trapped gases and dust. 
There is a large range of free parameters in the models. The spin state and specific
porous medium properties play a large role in the differences in activity between comets. The 
release of CO and subsequent permeability of the gas flow could inhibit emission,
especially if the local temperature (and pressure) gradients vary. This variation could be
a strong function of solar radiation heating and surface cooling, which are functions of the 
spin state and surface morphology. Since Rosetta, the latter has become an increasingly important factor \citep{thom15}.

Models for Chiron's distant activity are presented in \citet{pri95,cap00,fan97}. The first two include amorphous ice and rely on release of trapped CO, while the latter starts with a layered model of dust mantle, thick water ice layer and CO layer underneath. The amorphous ice models better reproduce the distant activity, although they are constrained by a lack of observations of CO and dust production. The latter is a convolution of gas fluxes, dust properties and coma behavior. Models dominated by crystalline water ice and CO/CO$_2$ ices are not successful \citep{pri95}. Instead, Chiron's activity appears to be best explained by a release of trapped volatile species in the water ice matrix. However, a more complete understanding of the outburst or quiescent nature of this activity would require more measurements of CO production rates, at different times. An optimal addition would be observations of other species (e.g., methanol, methane, CO$_2$ , etc.), which would enable more stringent constraints on the nucleus' composition ratios and thermal history. Radiogenic heating may be important for explaining Chiron's low-level activity that appears CO-depleted. 
Furthermore, if rings are confirmed around Chiron, then previously reported long-term changes in visual magnitude and infrared spectral ice features should be re-examined.

\section{Acknowledgements}

The authors thank the anonymous referee for useful comments and feedback, which improved the paper. This material is based upon work supported by (while MW was serving at) the National Science Foundation (NSF). 
MW also acknowledges support from NSF grants AST-1009933 and AST-1615917. Any opinion, findings, and conclusions or recommendations expressed in this material are those of the author(s) and do not necessarily reflect the views of the National Science Foundation. The SMT is operated by the Arizona Radio Observatory (ARO), the Steward Observatory, and the University of Arizona, with support through the NSF University Radio Observatories program (AST-1140030).

\vfil\eject
\bibliography{BigReferences}
\bibliographystyle{apj}


\begin{table}

\caption{Sublimation Temperatures of Cometary Species}

\begin{tabular}{lr}\hline

Species & Temperature$^a$ (K) \\\hline

N$_2$ & 22\\

CO & 25  \\

CH$_4$ & 31 \\

H$_2$S & 57 \\

C$_2$H$_2$ & 57 \\

H$_2$CO & 64 \\

CO$_2$ & 72  \\

HC$_3$N & 74 \\

NH$_3$ &78 \\

CS$_2$ & 78 \\

SO$_2$ &83 \\

CH$_3$CN & 91 \\

HCN & 95 \\

CH$_3$OH & 99 \\

H$_2$O & 152  \\

\hline

\tablenotetext{a}{\citep{yam85,sek96}}
\end{tabular} 

\label{tab:svolatile} 

\end{table}

\vfil\eject

\begin{table}

\caption{Representative Gaseous Production Rates in Three Comets Beyond 4 AU}

\begin{tabular}{llrrl}\hline

Comet & Molecule  & r (AU) & Q (number/s) &  Reference \\\hline

29P & CN &  5.8 & 8.0\ee{24} & \citep{coch91} \\

29P & CO & 5.7 - 6.2 & 1-7\ee{28} & various$^a$ \\

29P & H$_2$O & 6.2 & 6.3\ee{27} & \citep{oot12} \\

29P & CO$_2$ & 6.2 & $<$3.5e{26}& \citep{oot12} \\

29P & HCN & 6.3 & $<$ 1.4\ee{27} & \citep{pag13} \\

29P & CH$_3$OH & 6.3 & $<$9.3\ee{27} & \citep{pag13} \\

29P & CH$_4$ & 6.3 & $<$1.3\ee{27} & \citep{pag13} \\

Chiron & CN &   11.3 & 3.7\ee{25} & \citep{bus91} \\

Chiron & CO &  8.5 & 1.3\ee{28}& \citep{wom99}, recalculated \\


Chiron & H$_2$O (via OH)& 8.5 & $<$4\ee{30} & \citep{par97} \\


Chiron & H$_2$S & 8.5 & $<$2.1\ee{27} & \citep{rau97} \\

Chiron & H$_2$CO & 8.5 & $<$4.2\ee{26} & \citep{rau97} \\

Hale-Bopp & CN & 6.8 &  6\ee{25} & \citep{fitz96} \\

Hale-Bopp & CN & 9.8 & 0.9\ee{25} & \citep{rauer03} \\


Hale-Bopp & CO & 6.8 & 1.8\ee{28}& \citep{jew96} \\

Hale-Bopp & H$_2$O & 4.8 & 2.0\ee{28} & \citep{weaver97} \\

Hale-Bopp & CO$_2$ & 4.6 & 1.3\ee{28} & \citep{cro99} \\

Hale-Bopp & HCN & 4.8 & 2\ee{26} & \citep{jew96} \\

Hale-Bopp & CH$_3$OH & 5.0 & 3.3\ee{27} & \citep{wom97} \\

Hale-Bopp & H$_2$CO & 4.1 &5.8\ee{26}  & \citep{biv99} \\


Hale-Bopp & C$_3$ & 7.0 & $\leq$ 0.6\ee{25} & \citep{rauer03} \\
\hline

\tablenotetext{a}{The range of CO production rates typically observed is listed, taken from \citet{sen94,cro95,fes01,gun08,pag13}}
\end{tabular} 
\label{tab:volatile}
\end{table}
\vfil\eject

\begin{table}
\vtop to\vsize{
\caption{CN/CO production rate ratios in three distant comets}
\begin{tabular}{llll}\hline
Comet  & $r$ (AU) & Q(CN)/Q(CO)    & Reference \\\hline
Chiron & 11, 8.5 & 0.0025    & \cite{bus91,wom99} \\
29P    & 5.8 & 0.0003      & \cite{coch91,gun08}$^a$   \\
Hale-Bopp  &  6.8  & 0.0033   & \cite{fitz96,jew96} \\
Hale-Bopp   & 6.0  & 0.0040  & \cite{rau97,biv02} \\
\hline

\tablenotetext{a}{This ratio for 29P may typically be significantly lower, as CN is rarely observed in this comet.}
\end{tabular} 
\label{tab:cnco}
\vfil}
\end{table}

\vfil\eject

\begin{table}
\vtop to \vsize{
\caption{Ratios of CO$_2$/CO production rates in distant comets}
\begin{tabular}{llll}\hline
Comet & $r$ (AU) & Q(CO$_2$)/Q(CO)  & Reference \\\hline
29p & 6.2 & $<$0.0125  & \citet{oot12} \\
Hale-Bopp & 4.6 & 0.10  & \citet{cro99} \\
Hale-Bopp & 4.9 & 0.19 & \citet{cro99} \\
Hale-Bopp & 4.0 & 0.55 & \citet{weaver97} \\
Hale-Bopp & 3.89 & 0.10 &  \citet{cro99} \\
\hline
\end{tabular}
\label{tab:co2co}
\vfil}
\end{table}

\vfil\eject

\begin{table}

\caption{CO Production Rates in Comet 29P}

\begin{tabular}{rllll}\hline

Date  & r (AU) & Q$^a$ (\ee{28} molecules s$^{-1}$)  \\\hline
25-Feb-16	&	5.96	&	5.3	$\pm$	1.3	\\
26-Feb-16	&	5.96	&	3.8	$\pm$	0.9	\\
28-Feb-16	&	5.96	&	5.1	$\pm$	1.3	\\
29-Feb-16	&	5.96	&	2.5	$\pm$	0.6	\\
1-Mar-16	&	5.96	&	2.8	$\pm$	0.7	\\
3-Mar-16	&	5.96	&	2.6	$\pm$	0.7	\\
21-Mar-16	&	5.95	&	3.2	$\pm$	0.8	\\
28-Mar-16	&	5.95	&	2.6	$\pm$	0.7	\\
31-Mar-16	&	5.95	&	1.0	$\pm$	0.2	\\
9-Apr-16	&	5.95	&	0.9	$\pm$	0.2	\\
15-Apr-16	&	5.95	&	0.8	$\pm$	0.2	\\
24-Apr-16	&	5.94	&	1.0	$\pm$	0.2	\\
29-May-16	&	5.93	&	1.0	$\pm$	0.3	\\
\hline
\tablenotetext{a}{Production rates derived from observations of the CO J=2-1 line using the Arizona Radio Observatory 10-m
Submillimeter Telescope (SMT). Full details in Wierzchos et al., in preparation.}
\end{tabular}
\label{tab:29pqco}
\end{table}

\vfil\eject




\vfil\eject

\vfil\eject

\vfil\eject

\vfil\eject

\vfil\eject


\end{document}